\documentclass[twocolumn,showpacs,preprintnumbers,amsmath,amssymb]{revtex4} \usepackage{dcolumn}

\usepackage{graphicx} \usepackage{bm}

\newcommand{\hI}{\hat{I}}
\newcommand{\hK}{\hat{K}}
\newcommand{\hM}{\hat{M}}
\newcommand{\hU}{\hat{U}}
\newcommand{\hV}{\hat{V}}

\newcommand{\hb}{\hat{\beta}}
\newcommand{\hd}{\hat{\delta}}

\newcommand{\hH}{\hat{H}}

\begin{document}
\title{Non-locality of Foldy-Wouthuysen and related transformations for the Dirac equation}
\date{\today}
\author{Tomasz M. Rusin*}
\email{Tomasz.Rusin@centertel.pl}
\author{Wlodek Zawadzki\dag}
\affiliation{*Orange Customer Service sp. z o. o., ul. Twarda 18, 00-105 Warsaw, Poland\\
            \dag Institute of Physics, Polish Academy of Sciences, Al. Lotnik\'ow 32/46, 02-688 Warsaw, Poland}


\begin{abstract}
Non-localities of Foldy-Wouthuysen and related transformations, which are
used to separate positive and negative energy states in the Dirac
equation, are investigated. Second moments of functional kernels generated
by the transformations are calculated, the transformed functions and their
variances are computed. It is shown that all the transformed quantities
are smeared in the coordinate space by the amount comparable to the
Compton wavelength~$\lambda_c=\hbar/mc$. \end{abstract}

\maketitle

\section{Introduction}

The Dirac equation for relativistic electrons, in spite of its fundamental
importance in physics, is far from being thoroughly understood~\cite{Dirac1928}. One
of its special features is that, even in the absence of fields, there
exists a spectrum of positive and negative electron energies. This
peculiarity is a source of various problems. One of them is a phenomenon
of Zitterbewegung (trembling motion) which arises from an interference of
positive and negative energy states
\cite{Schrodinger1930,BjorkenBook,SakuraiBook,GreinerBook}.
There exist attempts to circumvent this duality.
One of them is the Foldy and Wouthuysen transformation (FWT) which, for the case of no
external fields, allows one
to break the Dirac equation into separate equations for positive and
negative energies~\cite{Foldy1950}. In their original paper, FW remarked that a
functional kernel, which transforms functions from the original
representation to the FW representation, is characterized by a
non-locality in coordinate space of the order of the Compton wavelength
$\lambda_c=\hbar/mc$. Rose in his book~\cite{RoseBook} put this statement on a quantitative
basis by showing that the second moment of the kernel is equal to
$(3/4)\lambda_c^2$ (see below). This result was not followed by other
investigations and it is by now not well known. Also, the moments are
often used in statistical physics but they are a rather unorthodox way to
characterize quantum mechanical properties.
It was shown later that the FW transformation is not unique. In other
words, there exist other transformations capable of separating positive
and negative energy states in the Dirac equation, both in the absence
of fields and in the presence of an external magnetic
field~\cite{Case1954,deVries1970,Tsai1973,Weaver1975}. In
the following we consider, in addition to the FWT, a two-step
transformation proposed by Moss and Okninski (MO, Ref.~\cite{Moss1976}).

Let us consider, as a matter of example, an average value of the
time-dependent velocity operator~$\hat{v}(t)$ in the Heisenberg picture,
\begin{equation} \label{Iv(t)}
\langle \Phi|\hat{v}(t)|\Phi\rangle=\langle\Phi\hU^{\dagger}|\hU\hat{v}(t)\hU^{\dagger}|\hU\Phi\rangle,
\end{equation}
where $\hat{\bm v}(t)=e^{i\hH_D t/\hbar}\hat{v}e^{-i\hH_D t/\hbar}$, $\hH_D$ is the Dirac Hamiltonian
and~$|\Phi\rangle$ is an arbitrary Dirac spinor.
According to the Dirac equation one has
$\hat{v}_j=c\hat{\alpha}_j$, where~$\hat{\alpha}_j$ ($j=x,y,z$) are the
standard Dirac~$4\times 4$ matrices.
As is well known,~$\hat{\alpha}_j$ do not commute with~$\hH_D$, so the velocity
depends on time also in
the absence of fields, which results in the Zitterbewegung mentioned
above. The operator~$\hU$ in Eq.~(\ref{Iv(t)}) stands for a unitary
transformation, for example the FWT. Equation~(\ref{Iv(t)}) expresses the
well known fact that a unitary transformation does not change an average
value. In our case it means that a transformation will not eliminate the
Zitterbewegung, which makes physical sense. We wrote down Eq.~(\ref{Iv(t)})
to illustrate symbolically that, if one transforms the operators:
$\tilde{\hat{v}}(t) = \hU\hat{v}(t)U^{\dagger}$, one should also transform the
functions: $\tilde{\Phi}=\hU \Phi$. Important operators transformed according to FWT are given
in the original FW paper and quoted in many textbooks. Here, we
concentrate on the transformed functions which were not analyzed in the literature.

Our work has two objectives. The first is to characterize in various ways the
non-locality of the Foldy-Wouthuysen transformation for the Dirac equation.
In doing this we continue the work of Rose but use means more typical for
quantum mechanics. The second objective is to investigate, as a matter of
example, another ''separating`` transformation and to compare its properties
to those of FWT. Such a comparison should give an idea of what one can
expect of various ''separating`` transformations.

Our subject is of relevance for two reasons. First, with ''the rise of
graphene`` there is nowadays a great deal of interest in
relativistic-type wave equations~\cite{Novoselov2004,Zawadzki2005,Zawadzki2011}.
Second, it is now possible to simulate the Dirac equation with the
trapped ions and cold atoms interacting with the laser radiation,
where one can tailor much more ''user
friendly`` values of the basic parameters~$mc^2$
and~$c$~\cite{Leibfried2003,Lamata2007,Johanning2009,Rusin2010}.
In fact, a proof-of-principle experiment simulating 1+1 Dirac equation and the
resulting Zitterbewegung was carried out by Gerritsma {\it et al.}~\cite{Gerritsma2010}.
If one deals with effective parameters by employing
narrow-gap semiconductors, the non-locality is determined by an ''effective``
Compton wavelength~$\lambda_Z=\hbar/m^*u$, where~$m^*$ is the effective electron mass
and~$u$ is the maximum electron velocity. In narrow-gap semiconductors one can deal with
$m^*~\simeq 10^{-2}m$ and $u\simeq 10^8$ cm/s, so that
the non-locality can be a few orders of magnitude larger
than that in a vacuum, see Ref.~\cite{Zawadzki2005}. The band
structure of graphene corresponds to the ''ultra-relativistic``
case (no gap) and a similar maximum velocity~$u$, so the non-locality may be even
larger.

\section{Transformation kernels and non-locality}
In this section we establish forms of Foldy-Wouthuysen and Moss-Okninski transformations
and define corresponding transformation kernels. Next we characterize nonlocal properties of the
kernels by calculating their moments.
The Dirac equation for a free relativistic electron is
$\hH_{\bm p}=c\hat{\bm \alpha}\hat{\bm p}+\hb mc^2$, where~$\hat{\bm p}$ is the momentum
and~$\hat{\bm \alpha}_j$ and~$\hb$ are~$4\times 4$ Dirac matrices in the standard notation.
The eigenenergies are~$\pm E_p$ with $E_p=\sqrt{m^2c^4+c^2p^2}$.

For the field-free case the momentum components are good quantum numbers $\hat{\bm p}\rightarrow {\bm p}$.
In the FW transformation an initial function~$|\Phi\rangle$ is
transformed to the new representation~$|\tilde{\Phi}\rangle$ with the use of the operator
$\hU_{FW}$ defined as
\begin{equation}
\label{U_FW}
 \hU_{FW} = \frac{E_p + \hb \hH_{\bm p}}{\sqrt{2E_p(E_p+mc^2)}}.
\end{equation}
There is $|\tilde{\Phi}\rangle = \hU_{FW}|\Phi\rangle$.
To find the MO transformation, we follow Moss and Okninski~\cite{Moss1976} and
introduce first the unitary operator
\begin{equation} \label{Up_V} \hV=\hd(\hd+\hb)/\sqrt{2},\end{equation}
which transforms the Dirac equation to the form
$\hH'_{\bm p}=c\hat{\bm \alpha}\hat{\bm p}+\delta mc^2$, in
which
\begin{equation}\hd=\hat{\alpha}_x\hat{\alpha}_y\hat{\alpha}_z\hb =
 \left(\begin{array}{cccc} 0 & 0 &-i & 0\\ 0 & 0 & 0 & -i\\
  i & 0 & 0 & 0 \\ 0 & i & 0 &0 \end{array}\right).\end{equation}
The Hamiltonian~$\hH'_{\bm p}$ has zeroes on the diagonal.
For a four-component function~$|\Psi\rangle$ there is $|\Psi'\rangle=\hV|\Psi\rangle$.
The two-step MO transformation is defined using the~$\hH'_{\bm p}$ Hamiltonian
in the following way
\begin{equation} \label{Up_MO}
 \hU_{MO} = \frac{1}{\sqrt{2}}\left(\hb + \frac{\hH'_{\bm p}}{E_p} \right).
\end{equation}

In the field-free case the operators~$\hU_{FW}$ and~$\hU_{MO}$ are number matrices.
Following Foldy and Wouthuysen we introduce a function transformation from the old to the new
representation $\langle {\bm r}|\tilde{\Phi}\rangle=\langle {\bm r}|\hU\Phi\rangle$. Inserting the
unity operators we obtain
\begin{eqnarray}\label{UK_Phi}
 \tilde{\Phi}({\bm r}) &=& \int\!\!\!\int\!\!\!\int \frac{ d^3{\bm p}d^3{\bm p}'d^3{\bm r}'}{(2\pi\hbar)^6}
  \langle{\bm r}|{\bm p}
  \rangle \langle {\bm p}|\hU|{\bm p}' \rangle \langle {\bm p}'|{\bm r}' \rangle \langle{\bm r}'|\Phi\rangle
   \nonumber \\
 &=& \int\!\!\int\!\!\frac{d^3{\bm p}d^3{\bm r}'}{(2\pi\hbar)^3}
     e^{i{\bm p}({\bm r}-{\bm r}')/\hbar} \hU_{\bm p} \Phi({\bm r}')
     \nonumber \\
 &=& \int \hK({\bm r},{\bm r}') \Phi({\bm r}') d^3 {\bm r}',
\end{eqnarray}
which defines the transformation kernel
\begin{equation} \label{UK_K}
 \hK({\bm r},{\bm r}') = \int \frac{e^{i{\bm p}({\bm r}-{\bm r}')/\hbar}}{(2\pi\hbar)^3} \hU_{\bm p} d^3 {\bm p}.
\end{equation}
In Eqs.~(\ref{UK_Phi}) and~(\ref{UK_K})~$\hU$ stands for each of the two operators defined above
and $\hU_{\bm p}=\langle {\bm p}|\hU|{\bm p}'\rangle\delta({\bm p}-{\bm p}')$.
The kernels are~$4\times 4$ matrices.

As mentioned above, the transformation kernels
have a nonlocal character. We also remarked that Rose~\cite{RoseBook} described
this feature quantitatively by
calculating the second moment of the FW kernel and showing that its smearing is
given by the Compton wavelength~$\lambda_c$. Below we evoke his calculations,
as it is not easily accessible by now, and compare his result with the corresponding
quantities for the MO transformation. The moments are rarely used to describe
quantum mechanical behavior, for this reason we do not limit our subsequent considerations
to the moments but also consider other quantities.

All the elements of matrices defined in Eq.~(\ref{UK_K}) are nonlocal functions of
${\bm R}={\bm r}-{\bm r}'$. To estimate the non-locality we calculate the zeroth and
second moments~$\hM^{(n)}$ of~$\hK_{FW}({\bm R})$ and~$\hK_{MO}({\bm R})$ matrices
\begin{equation}
 \hM^{(n)}_{ij} = \int \hK_{ij}({\bm R}) {\bm R}^n d^3{\bm R}.
\end{equation}
Let us consider first the~$(1,1)$ element of~$\hK_{FW}({\bm R})$.
For the zeroth moment there is
\begin{equation}
 \hM^{(0)}_{11} = \int d^3{\bm R} \int \frac{e^{i{\bm p}{\bm R}/\hbar}}{(2\pi\hbar)^3}
       \frac{E_p+mc^2}{\sqrt{2E_p(E_p+mc^2)}} d^3{\bm p}.
\end{equation}
Changing the order of integration one has
$\int e^{i{\bm p}{\bm R}/\hbar}d^3{\bm R} =(2\pi\hbar)^3\delta(\bm p)$. Then the integration
over~$d^3{\bm p}$ is trivial and one obtains~$ \hM^{(0)}_{11}=1$. The second
moment of the~$(1,1)$ element is
\begin{equation}
 \hM^{(2)}_{11} = \int d^3{\bm R} \int \frac{e^{i{\bm p}{\bm R}/\hbar}}{(2\pi\hbar)^3}
       \frac{E_p+mc^2}{\sqrt{2E_p(E_p+mc^2)}} {\bm R}^2 d^3{\bm p}.
\end{equation}
Integrating twice by parts one obtains
\begin{eqnarray}
 \hM^{(2)}_{11} &=& \int d^3{\bm R} \int \frac{e^{i{\bm p}{\bm R}/\hbar}}{(2\pi\hbar)^3}
      \nabla^2 \left[ \frac{-\hbar^2(E_p+mc^2)}{\sqrt{2E_p(E_p+mc^2)}} \right]
         d^3{\bm p} \nonumber \\
      &=& -\hbar^2 \nabla^2 \left. \left[ \frac{E_p+mc^2}{\sqrt{2E_p(E_p+mc^2)}}
       \right] \right|_{{\bm p} ={\bm 0}} = \frac{3}{4}\lambda_c^2.
\end{eqnarray}
In the same way one calculates moments of the remaining matrix elements.
We obtain $\hM^{(0)}_{FW}=\hI$ and $\hM^{(2)}_{FW}=(3/4)\lambda_c^2 \hI$, where~$\hI$ is the
identity matrix. These are the results of Rose~\cite{RoseBook}.

Thus the~$\hK_{FW}$ kernel is smeared in the ${\bm R}={\bm r} - {\bm r}'$ space by
the amount of~$\lambda_c/2$ in each direction. The kernel can be considered to be a transformed
Dirac delta function, cf. Eq.~(\ref{UK_Phi}). Hence, one can subscribe to the statement of
Foldy and Wouthuysen that ``a wave function which in the old representation corresponds to a state
in which the particle was localized at one point, passes over in the new representation
into a wave function which corresponds to the particle being spread out over a
finite region''. The physical meaning of this result is discussed in
Refs.~\cite{Foldy1950,RoseBook}.

We carry similar calculations for the Moss-Okninski transformation and calculate the zeroth
moment of the~$(1,3)$ element of~$\hU_{MO}$
\begin{equation}
 \hM^{(0)}_{13} = \int d^3{\bm R} \int \frac{e^{i{\bm p}{\bm R}/\hbar}}{(2\pi\hbar)^3}
       \frac{imc^2+cp_z}{\sqrt{2}E_p} d^3{\bm p} = \frac{i}{\sqrt{ 2}}.
\end{equation}
For the second moment of~$(1,3)$ element of~$\hU_{MO}$ we integrate twice by parts and obtain
\begin{eqnarray}
 \hM^{(2)}_{13} &=& \int d^3{\bm R} \int \frac{e^{i{\bm p}{\bm R}/\hbar}}{(2\pi\hbar)^3}
      \nabla^2 \left[\frac{imc^2+cp_z}{\sqrt{2}E_p} \right]
         d^3{\bm p} \nonumber \\
      &=& -\hbar^2 \nabla^2 \left. \left[\frac{imc^2+cp_z}{\sqrt{2}E_p}
       \right] \right|_{{\bm p} ={\bm 0}} = \frac{3i\lambda_c^2}{\sqrt{2}}.
\end{eqnarray}
Calculating the remaining moments one finally obtains
\begin{eqnarray}
\hM^{(0)}_{MO} &=& \frac{1}{\sqrt{2}} \left(\hb + \hd \right), \\
\hM^{(2)}_{MO} &=& 3\frac{\lambda_c^2}{\sqrt{2}}\ \hd.
\end{eqnarray}
Thus the kernels of both FW and MO transformations have finite non-localities of the
order of~$\lambda_c$, but they differ somewhat from each other.
In this sense the FW transformation is somewhat more ``compact'' than the MO
transformation.

\section{Transformation of functions}

We further investigate properties of FW and MO transformations by studying the transformed
functions. Let us consider an initial wave function $\Psi=f({\bm r})(1,0,0,0)^T$,
where~$f({\bm r})$ is normalized.
Then the transformed function for the MOT is
$\tilde{\Psi}'({\bm r})=\tilde{\Psi}'_1({\bm r}) + \tilde{\Psi}'_2({\bm r})$,
in which
\begin{eqnarray} \label{FK_Psi_1}
\tilde{\Psi}'_1({\bm r}) &=& \frac {1}{2} f({\bm r}) (1 , 0 ,-i ,0)^T, \\
\label{FK_Psi_2}
 \tilde{\Psi}'_2({\bm r}) &=&
 \frac {1}{2}\int \frac{d^3{\bm p} }{(2\pi\hbar)^3}\frac{e^{i{\bm p}{\bm r}/\hbar} f_{\bm p}}{ E_p}
   \left( \begin{array}{c} mc^2 + icp_z \\ icp^+\\ imc^2+cp_z\\ cp^+ \end{array} \right)\!\! ,
\end{eqnarray}
where
\begin{equation} \label{FK_fp}
 f_{\bm p} = \int e^{-i{\bm p}{\bm r}'/\hbar} f({\bm r}') d^3 {\bm r}'.
\end{equation}
The prime in~$\tilde{\Psi}'({\bm r})$ means that we transform the function
$\Psi'({\bm r})=\hV\Psi({\bm r})$, see Eq.~(\ref{Up_V}).

If the initial function $\Phi({\bm r})=f({\bm r})(1,0,0,0)^T$ is transformed according to FWT, there is
\begin{equation} \label{FK_Phi}
\tilde{\Phi}({\bm r})=
 \int \frac{d^3{\bm p} }{(2\pi\hbar)^3}\frac{e^{i{\bm p}{\bm r}/\hbar} f_{\bm p}}{N_p}
   \left( \begin{array}{c} E_p+ mc^2 \\ 0\\ -cp_z\\ -cp^+ \end{array} \right),
\end{equation}
where $N_p = \sqrt{2E_p(E_p+mc^2)}$. The asymmetry between two upper and two lower
components of~$\tilde{\Phi}({\bm r})$ arises from the asymmetry of
components in the initial wave packet.

Let us consider first the initial function in form of the delta function:
$\Psi({\bm r}) = \delta({\bm r})(1,0,0,0)^T$. Then~$f_{\bm p}\equiv 1$ and
the transformed function is obtained from Eqs.~(\ref{FK_Psi_1}) and~(\ref{FK_Psi_2}) in
terms of four integrals~$D_0$,~$D_x$,~$D_y$,~$D_z$
\begin{equation} \label{FK_DDelta}
\tilde{\Psi}'_{MO}({\bm r})= \frac{\delta({\bm r})}{2} \left( \begin{array}{c} 1 \\ 0 \\ -i \\ 0 \end{array} \right) +
  \left( \begin{array}{c} D_0+iD_z \\ iD_x-D_y\\ iD_0+D_z\\ D_x+iD_y \end{array} \right).
\end{equation}
The first integral is
\begin{equation} \label{FK_D0}
 D_0= \int \frac{d^3{\bm p} }{2(2\pi\hbar)^3}\frac{e^{i{\bm p}{\bm r}/\hbar} mc^2}{ E_p}=
      \frac{1}{(2\pi)^2\lambda_c^3}\frac{\lambda_c}{r}K_1(r/\lambda_c),
\end{equation}
where~$K_1(z)$ is the modified Bessel (MacDonald) function. To get Eq.~(\ref{FK_D0})
we used identities~(\ref{AK0}) and~(\ref{AK1R}) in Appendix. The second integral is
\begin{equation}
 D_z = \int \frac{d^3{\bm p} }{2(2\pi\hbar)^3}\frac{e^{i{\bm p}{\bm r}/\hbar} cp_z}{ E_p}.
\end{equation}
This integral is divergent, so we separate it into the divergent and convergent parts
and obtain
\begin{eqnarray} \label{FK_Dz}
 D_z &=& \int \frac{d^3{\bm p} e^{i{\bm p}{\bm r}/\hbar}}{2(2\pi\hbar)^3} {\rm sgn}(p_z)
        \nonumber \\
    &+& \int \frac{d^3{\bm p} e^{i{\bm p}{\bm r}/\hbar} }{2(2\pi\hbar)^3}
     \left[\frac{cp_z}{\sqrt{m^2c^4 +c^2p^2}} - {\rm sgn}(p_z) \right] \nonumber \\
     &=& \frac{2\lambda_c}{z}\delta({\bm \rho}) + {\cal A}({\bm r}/\lambda_c),
\end{eqnarray}
where~${\bm \rho}=(x,y)$.
We marked the integral in the second line of Eq.~(\ref{FK_Dz}) by~${\cal A}({\bm r}/\lambda_c)$.
This integral is convergent since the integrand has no singularities and for large~$p$
it decreases as~$p^{-2}$. The integral~${\cal A}({\bm r}/\lambda_c)$ can be expressed
in terms of the Anger functions and it decays exponentially with~$|{\bm r}|$ with a
characteristic length~$\lambda_c$. The integrals~$D_x$ and~$D_y$ can be obtained from~$D_z$
replacing~$z$ by~$x$ and~$y$. In conclusion, it can be seen that the delta
function subjected to the MO transformation becomes a
function {\it having a finite width of the order of~$\lambda_c$}.

Similar calculations can be performed for a function
$\Phi({\bm r}) = \delta({\bm r})(1,0,0,0)^T$ subjected to the FW transformation.
The transformed function is a combination of four integrals:
$\tilde{\Phi}=(B_0,0,-B_z,-B_x-iB_y)^T$, see Eq.~(\ref{FK_Phi}).
Setting~$f_{\bm p}\equiv 1$ and separating out the divergent part we obtain
\begin{eqnarray}
B_0 &=&
  \int \frac{d^3{\bm p} e^{i{\bm p}{\bm r}/\hbar}}{(2\pi\hbar)^3\sqrt{2}}\left\{\left[
     \frac{ (E_p + mc^2) }
      {\sqrt{E_p(E_p+mc^2)}} -1 \right] +1\right\} \nonumber \\
&=&  \frac{\delta({\bm r})}{\sqrt{2}} + \int \frac{d^3{\bm k} }{\sqrt{2}(2\pi)^3}
       \frac{e^{i{\bm k}{\bm r}}}{\sqrt{1+\lambda_c^2k^2}} G(k),
\end{eqnarray}
where $G(k)=\left(1 + \sqrt{1+1/\sqrt{1+\lambda_c^2k^2}}\right)^{-1}$.
We changed~${\bm p}$ to~$\hbar {\bm k}$.
The function~$G(k)$ varies slowly from $G(0)=1/(\sqrt{2}+1)\approx 0.414$ to~$G(\infty)=1/2$.
Thus we may approximate~$B_0$ by setting~$G(k)$ to have a constant value~$C_0$. This gives
\begin{equation}
B_0 \simeq \frac{\delta({\bm r})}{\sqrt{2}}
   +\frac{C_0}{\sqrt{2}(2\pi)^2\lambda_c^3} \frac{\lambda_c}{r} K_1(r/\lambda_c).
\end{equation}
This final result is very similar to that for the MOT, see Eqs.~(\ref{FK_DDelta}) and~(\ref{FK_D0}).
The integrals~$B_x$,~$B_y$ and~$B_z$ can be calculated
in a similar way to that described for the MO transformation.
Thus for both MOT and FWT the initial delta function transforms into
functions of finite width of the order of~$\lambda_c$.

In the above analysis with the initial delta function the
interpretation of results is somewhat difficult because of divergent and singular
final integrals. To avoid these problems we take the initial function in the
form of a Gaussian packet having a finite width~$d$
\begin{equation} \label{FK_Gauss}
 f({\bm r}) = \frac{\exp(-r^2/2d^2)}{\pi^{3/4}d^{3/2}}.
\end{equation}
The packet is normalized according to $\int |f({\bm r})|^2d^3{\bm r}=1$.
Now there is $f_{\bm p}=(2d\sqrt{\pi})^{3/2}\exp(-p^2d^2/2\hbar^2)$ and all the integrals in
Eqs.~(\ref{FK_Psi_2}) and~(\ref{FK_Phi}) are quickly convergent. The transformed
function is again expressed in terms of four integrals~$T_0$,~$T_x$,~$T_y$ and~$T_z$
\begin{equation} \label{FK_DGauss}
\tilde{\Psi}({\bm r})= \frac{f({\bm r})}{2} \left( \begin{array}{c} 1 \\ 0 \\ -i \\ 0 \end{array} \right) +
  \left( \begin{array}{c} T_0+iT_z \\ iT_x-T_y\\ iT_0+T_z\\ T_x+iT_y \end{array} \right).
\end{equation}
The first integral is [see Eq.~(\ref{FK_Psi_2})]
\begin{equation} \label{FK_T0}
 T_0 =\sqrt{2} mc^2 (d\sqrt{\pi})^{3/2} \int \frac{d^3{\bm p} }{(2\pi\hbar)^3}\frac{e^{i{\bm p}{\bm r}/\hbar}
   e^{-p^2d^2/2\hbar^2}}{\sqrt{ m^2c^4+c^2p^2}}.
\end{equation}
Integrating over the angular variables in the spherical coordinates we obtain
\begin{equation}
 T_0 =\frac{(d\sqrt{\pi})^{3/2}}{\sqrt{2}\pi^2} \int_{0}^{\infty}
       \frac{\sin(kr)}{kr}\frac{e^{-k^2d^2/2}k^2dk }{\sqrt{1+k^2\lambda_c^2}}.
\end{equation}
Applying the identity: $1/a=\int_{-\infty}^{\infty} \exp(-a\eta^2)d\eta/\sqrt{\pi}$,
integrating over~$k$ with the use of formula~(\ref{AIntSin}), we get
\begin{equation} \label{FK_T0eta}
 T_0 =\frac{(d\sqrt{\pi})^{3/2}}{\sqrt{2}\pi^2}\int_{-\infty}^{\infty}
      \frac{ e^{-\eta^2} e^{-r^2/(2d^2+4\eta^2\lambda_c^2)}}
     {4(d^2/2+\eta^2\lambda_c^2)^{3/2}} d\eta.
\end{equation}
The presence of~$e^{-\eta^2}$ term in Eq.~(\ref{FK_T0eta}) reduces the range of
integration to~$|\eta| \le 3$.
In the limit of~$d\gg \lambda_c$ we can neglect~$\eta^2\lambda_c^2$
as compared to~$d^2$ and obtain
\begin{equation} \label{FK_T0_larged}
 T_0 \simeq \frac{(d\sqrt{\pi})^{3/2}}{\sqrt{2}\pi^2} \frac{e^{-r^2/2d^2}}{4(d^2/2)^{3/2}}
       \int_{-\infty}^{\infty}\hspace*{-0.5em} e^{-\eta^2} d\eta = \frac{e^{-r^2/2d^2}}{2\pi^{3/4}d^{3/2}}.
\end{equation}
Thus, there is no widening of the wave packet in the limit of large widths~$d$, and the
transformed packet is almost identical to the initial one. In the opposite limit of
very narrow packets:~$d \ll \lambda_c$,
we may neglect in Eq.~(\ref{FK_T0eta}) the~$d^2$ term under the integral sign. After the
substitution~$\chi=\eta^2$ one gets
\begin{equation}
 T_0 \simeq \frac{(d\sqrt{\pi})^{3/2}}{\sqrt{2}\pi^2\lambda_c^3} \int_{0}^{\infty}
      \frac{e^{-\chi- (r/\lambda_c)^2/(4\chi)} }{4\chi^2}d\chi.
\end{equation}
Using identities~(\ref{AIntD0}) in Appendix and~$K_{-\nu}(z)=K_{\nu}(z)$
we finally obtain
\begin{equation} \label{FK_T0_smalld}
 T_0 \simeq \frac{d^{3/2}}{\sqrt{2}\pi^{5/4}\lambda_c^3} \frac{\lambda_c}{r} K_1(r/\lambda_c).
\end{equation}
Thus in the limit of narrow packets the transformed function
acquires a width of the order of~$\lambda_c \gg d$.
For all values of~$d$ the width of the
transformed function is larger than the width of the initial packet.

For~$T_z$ integrals in Eq.~(\ref{FK_DGauss}) we have
\begin{eqnarray} \label{FK_Tz}
 T_z &=&\sqrt{2} (d\sqrt{\pi})^{3/2} \int \frac{d^3{\bm p} }{(2\pi\hbar)^3}\frac{e^{i{\bm p}{\bm r}/\hbar}
   e^{-p^2d^2/2\hbar^2} cp_z}{\sqrt{ m^2c^4+c^2p^2}} \nonumber \\
    &=& -i\lambda_c\frac{\partial T_0}{\partial z}.
\end{eqnarray}
Integrals~$T_x$ and~$T_y$ are obtained the same way.
They have similar properties to~$T_0$ integral.

It is seen that~$T_0$ of Eq.~(\ref{FK_T0_smalld}) differs from~$D_0$ of Eq.~(\ref{FK_D0})
by a factor of~$(2\sqrt{\pi}d)^{3/2}$. This is due to a different normalizations of the
packet~$f({\bm r})$ and the delta function. If one normalizes the packet according to
$\int f_{\delta}({\bm r})d^3{\bm r}=1$ the results of Eqs.~(\ref{FK_T0_smalld}) and~(\ref{FK_D0})
become the same.

For the FW transformation, a function~$\Phi({\bm r})=f({\bm r})(1,0,0,0)^T$
is transformed to $\tilde{\Psi}=(S_0,0,-S_z,-S_x-iS_y)^T$, where~$S_0$,~$S_x$,~$S_y$ and~$S_z$
are integrals defined below. There is [see Eq.~(\ref{FK_Phi})]
\begin{eqnarray} \label{FK_S0}
S_0 &=& \int \frac{d^3{\bm p} }{(2\pi\hbar)^3}\frac{e^{i{\bm p}{\bm r}/\hbar}
     f_{\bm p}(E_p+ mc^2)}{N_p} = \frac{(d\sqrt{\pi})^{3/2}}{\pi^2r}\! \times
      \nonumber \\
    && \int_{0}^{\infty} k\sin(kr)e^{-k^2d^2/2}\sqrt{1+\frac{1}{E_k}}\ dk,
\end{eqnarray}
where $E_k=\sqrt{1+\lambda_c^2k^2}$ and~$f_{\bm p}$ is defined in Eq.~(\ref{FK_fp}).
The integral~$S_x$ is
\begin{equation} \label{FK_Sx}
S_x = \int \frac{d^3{\bm p} }{(2\pi\hbar)^3}\frac{e^{i{\bm p}{\bm r}/\hbar}
     f_{\bm p}cp_x}{N_p},
\end{equation}
and similarly for~$S_y$ and~$S_z$. Integrals~$S_x$,~$S_y$ and~$S_z$
can be obtained as partial derivatives of an auxiliary integral
\begin{equation} \label{FK_Saux}
S_{aux} =\int \frac{d^3{\bm p}}{(2\pi\hbar)^3}\frac{e^{i{\bm p}{\bm r}/\hbar}
     f_{\bm p}}{N_p}
\end{equation}
with respect to~$x$,~$y$ and~$z$, respectively.

\begin{figure}
\includegraphics[width=8.0cm,height=8.0cm]{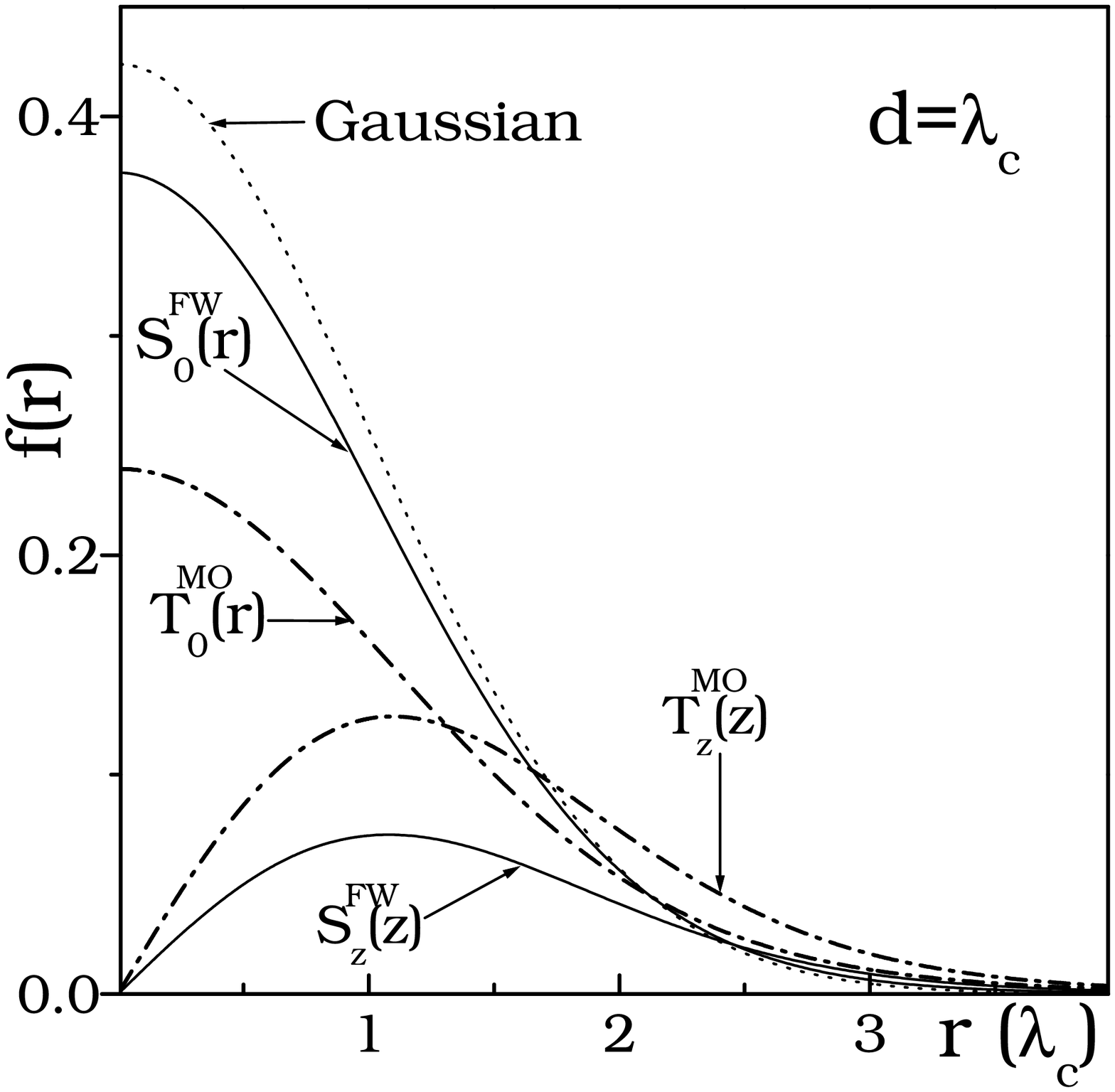}
\caption{Elements in the first component of the transformed function~$\tilde{\Psi}({\bm r})$
         (MOT) and~$\tilde{\Phi}({\bm r})$ (FWT) calculated for initial packet width~$d=\lambda_c$.
         Dotted line~$-$~the initial Gaussian wave packet.} \label{Fig1Fun}
\end{figure}

In Figure~\ref{Fig1Fun} we plot various elements of the transformed wave functions:
the integrals~$2T_0(r)$,~$2T_z(z)$,~$S_0(r)$ and~$S_z(z)$ calculated for~$d=\lambda_c$.
Solid lines represent~$S_0(r)$ and~$S_z(z)$ integrals, while dashed lines represent~$2T_0(r)$
and~$2T_z(z)$. The factor~$2$ in front of~$2T_0(r)$ and~$2T_z(z)$ integrals
is chosen to cancel out the factor~$1/2$ appearing in Eq.~(\ref{FK_Psi_2}).
The terms~$2T_z(z)$ and~$2S_z(z)$ are related to~$p_z$ terms in
Eqs.~(\ref{FK_Psi_2}) and~(\ref{FK_Phi}).

The integral~$S_0(r)$ is very similar the initial Gaussian function. For the MOT
the difference between the initial and the transformed functions is more pronounced.
For larger~$d$, the integrals~$S_0(r)$ and~$T_0(r)$ tend to the initial Gaussian function,
so almost no widening occurs. This agrees with the the large-$d$ approximation for~$T_0(r)$
in Eq.~(\ref{FK_T0_larged}).

\section{Variances}

Finally, we estimate the spatial extent of functions subjected to the MO and FW transformations
by calculating their normalized variances. For the MOT we define
\begin{equation} \label{V_Def}
 V_{MO} = \frac{\langle \tilde{\Psi}'|{\bm r}^2|\tilde{\Psi}'\rangle}{\langle \tilde{\Psi}'|\tilde{\Psi}'\rangle},
\end{equation}
and similarly for the FWT. Functions~$\tilde{\Psi}'({\bm r})$ and~$\tilde{\Phi}'({\bm r})$
are normalized to unity. We assume the initial function in the form of a Gaussian packet given in Eq.~(\ref{FK_Gauss}).
Since $\tilde{\Psi}'({\bm r})=\tilde{\Psi}'_1({\bm r})+\tilde{\Psi}'_2({\bm r})$,
see Eqs.~(\ref{FK_Psi_1}) and~(\ref{FK_Psi_2}),
it is convenient to introduce averages over the first and second parts
of~$\tilde{\Psi}'({\bm r})$. We define
$\langle r^n\rangle_{ab}= \langle \tilde{\Psi}'_a|{\bm r}^n|\tilde{\Psi}'_b\rangle$
with~$a,b=\{1,2\}$. Then
\begin{equation}
V_{MO} = \frac{\langle r^2\rangle_{11}+\langle r^2\rangle_{12}+\langle r^2\rangle_{21}+\langle r^2\rangle_{22}}
         {\langle r^0\rangle_{11}+\langle r^0\rangle_{12}+\langle r^0\rangle_{21}+\langle r^0\rangle_{22}}.
\end{equation}
There is~$\langle r^0\rangle_{11}=1/2$ and~$\langle r^2\rangle_{11}=3d^2/4$.
We consider now~$\langle r^n\rangle_{12}$ terms. Introducing
\begin{equation}
g_{\bm p}^{(n)} = \int e^{-i{\bm p}{\bm r}/\hbar} f({\bm r}) {\bm r}^n d^3{\bm r},
\end{equation}
we have
\begin{equation} \label{V_rn12}
 \langle r^n\rangle_{12}=\!\!\int\ \frac{d^3{\bm p}}{(2\pi\hbar)^3}
   \frac{g_{\bm p}^{(n)*}f_{\bm p}}{4E_p}
     \left( \begin{array}{c} 1 \\ 0\\ -i\\ 0 \end{array} \right)^{\!\!\dagger} \hspace*{-0.5em}
     \left( \begin{array}{c} mc^2 + icp_z \\ icp^+\\ imc^2+cp_z\\ cp^+ \end{array} \right)=0.
\end{equation}
The dagger denotes Hermitian conjugate of the vector and the star its complex conjugate.
In the above equation the terms proportional to~$mc^2$ cancel out. The terms
including~$p_z$ are odd functions of~$p_z$, so they vanish after the integration.
For the same reasons there is~$\langle r^n\rangle_{21}=0$.

To calculate~$\langle r^n\rangle_{22}$ we first introduce the Fourier transform
of~$\tilde{\Psi}'_2(\bm r)$
\begin{equation}
 \tilde{\Psi}'_{2{\bm p}}=\int e^{-i{\bm p}{\bm r}} \tilde{\Psi}'_2(\bm r) d^3{\bm r} =
        \frac{f_{\bm p}}{E_p} \vec{w}_{\bm p},
\end{equation}
where~$\vec{w}_{\bm p}$ is the column in Eq.~(\ref{FK_Psi_2}) or the second column
in Eq.~(\ref{V_rn12}). Then
\begin{eqnarray}
 \langle r^0\rangle_{22} &=&\int \frac{d^3{\bm r}d^3{\bm p}d^3{\bm p}'}{4(2\pi\hbar)^6}
    e^{i{\bm r}({\bm p}-{\bm p}')/\hbar}\tilde{\Psi}_{2{\bm p}'}^{,\dagger} \tilde{\Psi}'_{2{\bm p}}
     \nonumber \\
     &=& \int\frac{d^3{\bm p}}{4(2\pi\hbar)^3} \frac{|f_{\bm p}|^2}{E_p^2}
       \vec{w}_{\bm p}^{\dagger}\vec{w}_{\bm p} = \frac{1}{2}.
\end{eqnarray}
Since $\langle r^0\rangle_{11}=\langle r^0\rangle_{22}=1/2$ and
$\langle r^0\rangle_{12}=\langle r^0\rangle_{21}=0$, the function~$\tilde{\Psi}'({\bm r})$
is really normalized to unity.

Now we calculate~$\langle r^2\rangle_{22}$
\begin{equation}
 \langle r^2\rangle_{22}
  =\int \frac{d^3{\bm r}d^3{\bm p}d^3{\bm p}'}{4(2\pi\hbar)^6}
    e^{i{\bm r}({\bm p}-{\bm p}')/\hbar} {\bm r}^2\tilde{\Psi}_{2{\bm p}'}^{,\dagger}
   \tilde{\Psi}'_{2{\bm p}}.
\end{equation}
Integrating twice by parts over~$d^3{\bm p}$ and then integrating over~$d^3{\bm r}$ and~$d^3{\bm p}'$ one obtains
\begin{eqnarray}
 \langle r^2\rangle_{22}
  &=&-\hbar^2\int \frac{d^3{\bm r}d^3{\bm p}d^3{\bm p}'}{4(2\pi\hbar)^6}
    e^{i{\bm r}({\bm p}-{\bm p}')/\hbar} \tilde{\Psi}_{2{\bm p}'}^{,\dagger}
     \nabla^2_{\bm p}\tilde{\Psi}'_{2{\bm p}}
     \nonumber \\
     &=&-\hbar^2 \int\frac{d^3{\bm p}}{4(2\pi\hbar)^3}
      \left(\frac{f_{\bm p}\vec{\bm w}_{\bm p}}{E_p}\right)^{\dagger}
        \nabla^2_{\bm p} \left( \frac{f_{\bm p}\vec{\bm w}_{\bm p}}{E_p} \right).
\end{eqnarray}
After some manipulation we find
\begin{equation} \label{VA_r2_22}
\langle r^2\rangle_{22}= \frac{11}{4}d^2 - 2d^2\frac{\bar{d}}{\sqrt{\pi}}[A_{1}^0+A_{2}^0],
\end{equation}
where~$\bar{d}=d/\lambda_c$, and
\begin{equation} \label{V_DefA}
A_{\nu}^{\mu} = \int_{0}^{\infty} \frac{\exp(-t^2\bar{d}^2)\ t^{\mu}dt}{(1+t^2)^{\nu}}.
\end{equation}
For integer values of~$\nu$ and integer or half-integer values of~$\mu$ the integrals~$A_{\nu}^{\mu}$
can be expressed in terms of modified Bessel and error functions, see Appendix.

For the FW transformation we have similarly
\begin{equation}
V_{FW} = \frac{\langle \tilde{\Phi}|{\bm r}^2|\tilde{\Phi}\rangle}{\langle \tilde{\Phi}|\tilde{\Phi}\rangle}
       = \frac{\langle r^2\rangle_{FW}}{\langle r^0\rangle_{FW}},
\end{equation}
where
\begin{eqnarray}
 \langle r^0\rangle_{FW} &=&\int \frac{d^3{\bm r}d^3{\bm p}d^3{\bm p}'}{(2\pi\hbar)^6}
    e^{i{\bm r}({\bm p}-{\bm p}')/\hbar}\tilde{\Phi}_{{\bm p}'}^{\dagger} \tilde{\Phi}_{{\bm p}}=1, \\
 \langle r^2\rangle_{FW}
  &=&\int \frac{d^3{\bm r}d^3{\bm p}d^3{\bm p}'}{(2\pi\hbar)^6}
    e^{i{\bm r}({\bm p}-{\bm p}')/\hbar} {\bm r}^2\tilde{\Phi}_{{\bm p}'}^{\dagger} \tilde{\Phi}_{\bm p}
     \nonumber \\
     &=&-\hbar^2 \int\frac{d^3{\bm p}}{(2\pi\hbar)^3}
      \tilde{\Phi}_{\bm p}^{\dagger} \nabla^2_{\bm p} \tilde{\Phi}_{\bm p},
\end{eqnarray}
in which $\tilde{\Phi}_{{\bm p}}=\int e^{-i{\bm p}{\bm r}} \tilde{\Phi}(\bm r) d^3{\bm r}$.
After some algebra we obtain
\begin{equation} \label{VA_r2_FW}
 V_{FW} = \frac{7}{2}d^2 + d^2\frac{\bar{d}}{\sqrt{\pi}}
   [A_1^0 - A_2^0 - 4A_{1/2}^0 ].
\end{equation}

\begin{figure}
\includegraphics[width=8.0cm,height=8.0cm]{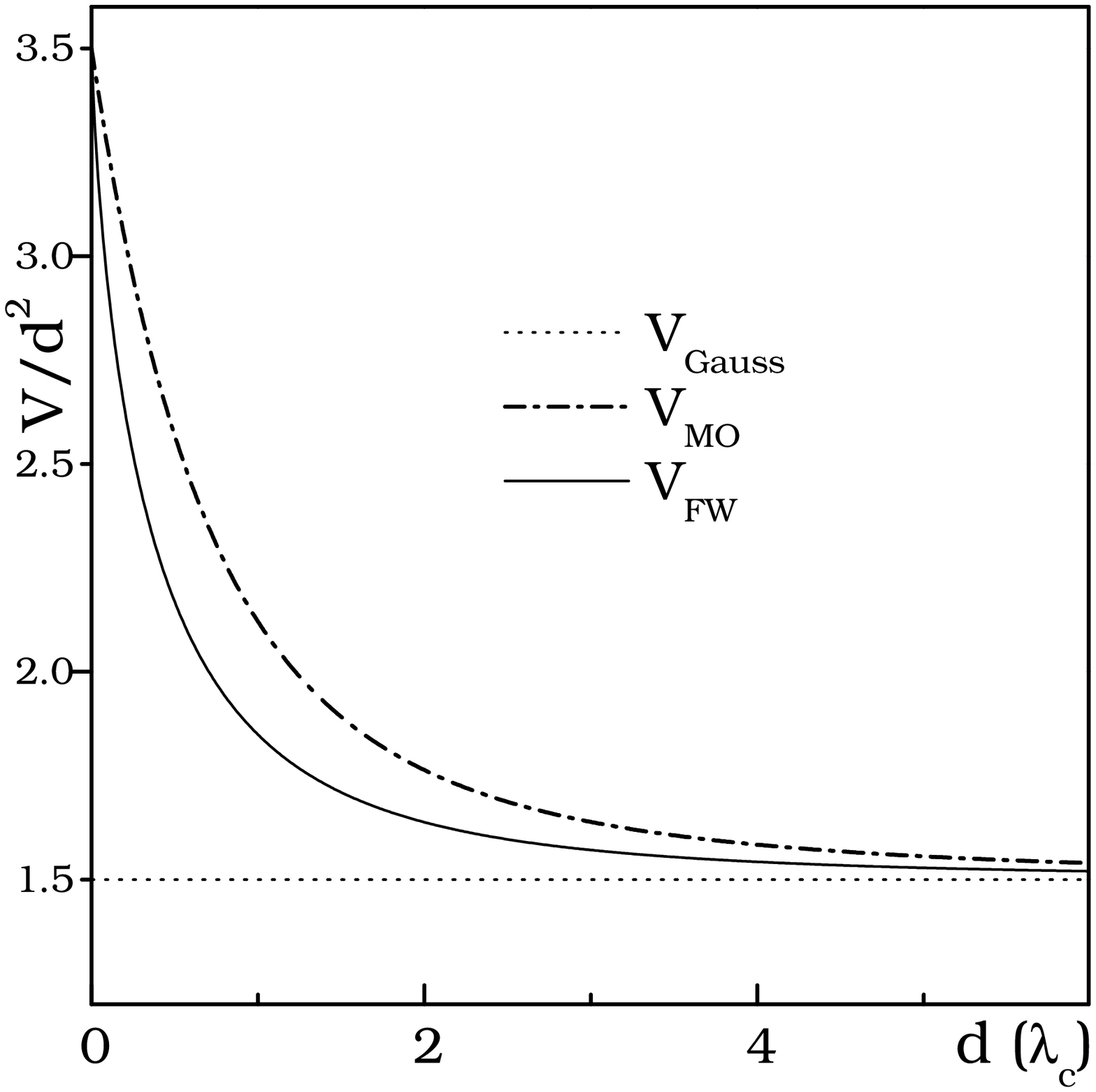}
\caption{Normalized variances of the transformed functions for the MOT and FWT calculated for various
         initial packet width~$d$.
         Solid line~$-$~FWT, dashed line~$-$~MOT.
         Dotted horizontal line~$-$~normalized variance of the initial Gaussian wave packet.
         The MOT and FWT variances has the small-$d$ limit~$7d^2/2$, while for large~$d$ they
         reach the Gaussian limit~$3d^2/2$.} \label{Fig2V}
\end{figure}

In Figure~\ref{Fig2V} we plot calculated variances of the transformed functions divided
by~$d^2$. For the initial Gaussian packet the variance is~$V_G=3d^2/2$ and it is
indicated by the horizontal dotted line. The other two variances
[MO-transformed function~$\tilde{\Psi}'({\bm r})$ and FW-transformed
function~$\tilde{\Phi}({\bm r})$] have very similar
dependencies on the packet width~$d$. Both variances have the zero-width
limit~$V_{d\rightarrow 0}=7d^2/2$ and they decrease with increasing~$d$
to the Gaussian limit~$V_G$.

Small-$d$ and large-$d$ limits of the variances can be obtained analytically using
Eqs.~(\ref{VA_r2_22}) and~(\ref{VA_r2_FW}).
Consider first the variance for~$\langle r^2\rangle_{22}$ from Eq.~(\ref{VA_r2_22}).
For small~$\bar{d}$ the second term in Eq.~(\ref{VA_r2_22}) tends to zero (see Appendix), and
there is~$\langle r^2\rangle_{22}\simeq 11d^2/4$. The first term is~$\langle r^2\rangle_{11}=3d^2/4$
and the total variance is~$V_{MO}\simeq (11d^2/4+3d^2/4)=7d^2/2$.
Similarly, for the FW transformation at small~${\bar{d}}$ the second term in Eq.~(\ref{VA_r2_FW})
vanishes and the total variance is again~$7d^2/2$.

The large-$\bar{d}$ limits of~$A_{\nu}^{\mu}$
in Eqs.~(\ref{VA_r2_22}) and~(\ref{VA_r2_FW}) are calculated in Appendix.
Applying these results to Eqs.~(\ref{VA_r2_22}) and~(\ref{VA_r2_FW})
we find that the large-$d$ limits of~$V_{MO}$ and~$V_{FW}$ are~$3d^2/2$,
i.e. they are equal to the variance of the initial Gaussian function.
Thus the widening of the transformed functions is pronounced for~$d<\lambda_c$ and
it is small for~$d\gg\lambda_c$, see Figure~\ref{Fig2V}. It can be seen that, again,
the FW transformation is more ''compact`` than the MO transformation because for
a given packet width~$d$ there is always~$V_{FW}< V_{MO}$.

\section{Conclusions and summary}

As mentioned in the Introduction, Rose~\cite{RoseBook} demonstrated that a non-locality
of the functional kernel generated by the Foldy-Wouthuysen transformation
is extended in the coordinate space over the Compton wavelength~$\lambda_c$.
We generalize this result to the transformed functions and another
transformation separating positive and negative energy states in the Dirac
equation (Moss-Okninski transformation). In particular, we show that the
delta function~$\delta({\bm r})$ is transformed into a nonlocal function smeared over
the distance~$\lambda_c$. Second-order variances are used to obtain similar
results for narrow Gaussian wave packets indicating that the non-locality
of the order of~$\lambda_c$ is a general property of the transformed functions.
Our results strongly suggest that all ''separating`` transformations for
the energies generate coordinate non-localities of this order.
We emphasize that for relativistic-type equations, appearing either in
narrow-gap semiconductors or in simulations, the non-localities can extend over
hundreds of angstroms.

\appendix
\section{}
We first quote formulas for the integrals appearing in the text.
Let~$R=\sqrt{a^2+b^2}$. Then
\begin{eqnarray}
\int_{-\infty}^{\infty}\frac{\exp(itz)dt}{\sqrt{t^2+a^2}}&=&2K_0(az), \label{AK0} \\
\int_0^{\infty}J_0(bt)K_0(a\sqrt{t^2+1})tdt&=& K_1(R)/R,              \label{AK1R}\\
\int_{0}^{\infty}k\sin(kr)\exp(-k^2p^2)dk&=&\frac{\sqrt{\pi}r}{4p^3}e^{-r^2/4p^2}, \label{AIntSin} \\
\int_0^{\infty} \chi^{\nu-1}\exp\left[-\chi-\frac{\mu^2}{4\chi}\right]d\chi
  &=&2\left(\frac{\mu}{2}\right)^{\nu}K_{-\nu}(\mu), \ \ \ \label{AIntD0}
\end{eqnarray}
where~$K_0(\xi)$,~$K_1(\xi)$ and~$K_{\nu}(\xi)$ are the modified Bessel (MacDonald) functions.
Next we collect formulas for~$A_{\nu}^{\mu}$ integrals.
Let~$D=\bar{d}^2/2$ and~${\rm Erf}(\xi)$ be the error function. Then
\begin{eqnarray}
A_{1/2}^0&=&\frac{1}{2}\exp(D)K_0(D), \\
A_1^0&=&\frac{\pi \exp(\bar{d}^2)[1-{\rm Erf}(\bar{d})]}{2}, \\
A_2^0&=&\frac{\pi \exp(\bar{d}^2)(2\bar{d}^2-1)[{\rm Erf}(\bar{d})-1]}{4}
      +\frac{\sqrt{\pi}\bar{d}}{2}. \ \ \ \ \ \
\end{eqnarray}
For~$\bar{d}\rightarrow 0$ there is
$\bar{d} A_{1/2}^0\simeq 0$, because in this limit there is $K_0(\bar{d})\simeq \ln(\bar{d})$~\cite{GradshteinBook}.
Also, $\bar{d}A_1^0\simeq 0$ and~$\bar{d}A_2^0\simeq 0$ because~${\rm Erf}(0)=0$.
Using large-$x$ expansions for the Bessel and error functions we obtain
\begin{eqnarray}
\lim_{\bar{d}\rightarrow \infty} \frac{\bar{d}}{\sqrt{\pi}} A_{1/2}^0&=&\frac{1}{2},\\
\lim_{\bar{d}\rightarrow \infty} \frac{\bar{d}}{\sqrt{\pi}} A_1^0    &=&\frac{1}{2},\\
\lim_{\bar{d}\rightarrow \infty} \frac{\bar{d}}{\sqrt{\pi}} A_2^0    &=&\frac{1}{2}.
\end{eqnarray}


\begin{thebibliography}{99}
\bibitem{Dirac1928}       P. A. M. Dirac, Proc. R. Soc. A {\bf 117}, 610 (1928).
\bibitem{Schrodinger1930} E. Schrodinger, Sitzungsber. Preuss. Akad.
                          Wiss. Phys. Math. Kl. {\bf 24}, 418 (1930).
                          Schrodinger's derivation is reproduced in
                          A. O. Barut and A. J. Bracken, Phys. Rev. D {\bf 23}, 2454 (1981).
\bibitem{BjorkenBook}     J. D. Bjorken and S. D. Drell, {\it Relativistic Quantum Mechanics} (McGraw-Hill, New York,1964).
\bibitem{SakuraiBook}     J. J. Sakurai {\it Modern Quantum Mechanics} (Addison-Wesley, New York, 1987).
\bibitem{GreinerBook}     W. Greiner {\it Relativistic Quantum Mechanics} (Springer, Berlin, 1994).
\bibitem{Foldy1950}       L. L. Foldy and S. A. Wouthuysen, Phys. Rev. {\bf 78} 29 (1950).
\bibitem{RoseBook}        M. E. Rose {\it Relativistic Electron Theory} (Wiley, New York, 1961).
\bibitem{Case1954}        K. M. Case, Phys. Rev. {\bf 95} 1323 (1954).
\bibitem{deVries1970}     E. de Vries, Fortschritte der Physik {\bf 18} 149, (1970).
\bibitem{Tsai1973}        W. Y. Tsai, Phys. Rev. D {\bf 7} 1945 (1973).
\bibitem{Weaver1975}      D. L. Weaver, Phys. Rev. D {\bf 12} 4001 (1975).
\bibitem{Moss1976}        R. E. Moss and A. Okninski, Phys. Rev. D {\bf 14}, 3358 (1976).
\bibitem{Novoselov2004}   K. S. Novoselov, A. K. Geim, S. V. Morozov, D. Jiang, Y Zhang, S. V. Dubonos, I. V. Grigorieva and A. A. Firsov,
                          Science {\bf 306}, 666 (2004).
\bibitem{Zawadzki2005}    W. Zawadzki, Phys. Rev. B {\bf 72}, 085217 (2005).
\bibitem{Zawadzki2011}    W. Zawadzki and T. M. Rusin, J. Phys. Cond. Matt. {\bf 23}, 143201 (2011).
\bibitem{Leibfried2003}   D. Leibfried, R. Blatt, C. Monroe, and D. Wineland, Rev. Mod. Phys. {\bf 75}, 281 (2003).
\bibitem{Lamata2007}      L. Lamata, J. Leon, T. Schatz, and E. Solano, Phys. Rev. Lett. {\bf 98}, 253005 (2007).
\bibitem{Johanning2009}   M. Johanning, A. F. Varron, and C. Wunderlich, J. Phys. B {\bf 42}, 154009 (2009).
\bibitem{Rusin2010}       T. M. Rusin and W Zawadzki, Phys. Rev. D {\bf 82} 125031 (2010).
\bibitem{Gerritsma2010}   R. Gerritsma, G. Kirchmair, F. Zahringer, E. Solano, R. Blatt and C. F. Roos,
                          Nature {\bf 463} 68 (2010).
\bibitem{GradshteinBook}  I. S. Gradshtein and I. M. Ryzhik 2007 {\it Table of Integrals, Series, and Products
                          (Ed. A Jeffrey and D Zwillinger 7th edition)}(Academic Press, New York, 2007).
\end{thebibliography}
\end{document}